\pdfoutput=1

\documentclass[11pt]{article}

\usepackage[]{acl}

\usepackage{times}
\usepackage{latexsym}

\usepackage[T1]{fontenc}

\usepackage[utf8]{inputenc}

\usepackage{microtype}

\usepackage{inconsolata}

\usepackage{graphicx}

\usepackage{times}
\usepackage{soul}
\usepackage{url}

\usepackage{amsmath}
\usepackage{amsthm}
\usepackage{booktabs}
\usepackage{algorithm}
\usepackage{algorithmic}
\usepackage[switch]{lineno}
\usepackage{graphicx}
\usepackage{multirow}
\usepackage{subcaption}
\usepackage{float}
\usepackage{balance}

\title{SymbioticRAG: Enhancing Document Intelligence Through Human-LLM Symbiotic Collaboration}

\author{
Qiang Sun\textsuperscript{1}\thanks{\ \ Corresponding author.} \quad 
Tingting Bi\textsuperscript{1,3} \quad 
Sirui Li\textsuperscript{2} \\
\textbf{Eun-Jung Holden}\textsuperscript{3} \quad 
\textbf{Paul Duuring}\textsuperscript{1,4} \quad 
\textbf{Kai Niu}\textsuperscript{1} \quad 
\textbf{Wei Liu}\textsuperscript{1}\footnotemark[1] \\
\textsuperscript{1}The University of Western Australia \quad 
\textsuperscript{2}Murdoch University \\
\textsuperscript{3}The University of Melbourne \quad 
\textsuperscript{4}Geological Survey of Western Australia \\
\texttt{\{pascal.sun,kai.niu\}@research.uwa.edu.au} \\
\texttt{sirui.li@murdoch.edu.au, \{tingting.bi,eunjung.holden\}@unimelb.edu.au} \\
\texttt{paul.duuring@demirs.wa.gov.au, wei.liu@uwa.edu.au}
}

\begin{document}

\maketitle

\begin{abstract}
We present \textbf{SymbioticRAG}, a novel framework that fundamentally reimagines Retrieval-Augmented Generation~(RAG) systems by establishing a bidirectional learning relationship between humans and machines. Our approach addresses two critical challenges in current RAG systems: the inherently human-centered nature of relevance determination and users' progression from "unconscious incompetence" in query formulation. SymbioticRAG introduces a two-tier solution where Level 1 enables direct human curation of retrieved content through interactive source document exploration, while Level 2 aims to build personalized retrieval models based on captured user interactions. We implement Level 1 through three key components: (1)~a comprehensive document processing pipeline with specialized models for layout detection, OCR, and extraction of tables, formulas, and figures; (2)~an extensible retriever module supporting multiple retrieval strategies; and (3)~an interactive interface that facilitates both user engagement and interaction data logging. We experiment Level 2 implementation via a retriever strategy incorporated LLM summarized user intention from user interaction logs. To maintain high-quality data preparation, we develop a human-on-the-loop validation interface that improves pipeline output while advancing research in specialized extraction tasks. Evaluation across three scenarios (literature review, geological exploration, and education) demonstrates significant improvements in retrieval relevance and user satisfaction compared to traditional RAG approaches. To facilitate broader research and further advancement of SymbioticRAG Level 2 implementation, our system is freely available via \url{https://app.ai4wa.com}.
\end{abstract}

\section{Introduction}\label{sec:intro}

Recent advances in large language models (LLMs) challenge the traditional search engines' information seeking convention by allowing users to pose natural, conversational queries that yield synthesized answers without explicitly retrieving source documents for further manual processing~\cite{zhu2023large}. 
Although this paradigm shift can be remarkably convenient for general inquiries, it runs into significant limitations when handling domain-specific or proprietary content that may not be present in the pre-trained language models. 
In such scenarios, relying solely on pre-trained models can exacerbate ``hallucinations'', where LLMs confidently present fabrications or partial truths~\cite{llmhallucination}.

\begin{figure*}[hbt]
\vspace{-1em}
\centering
\includegraphics[width=0.8\linewidth]{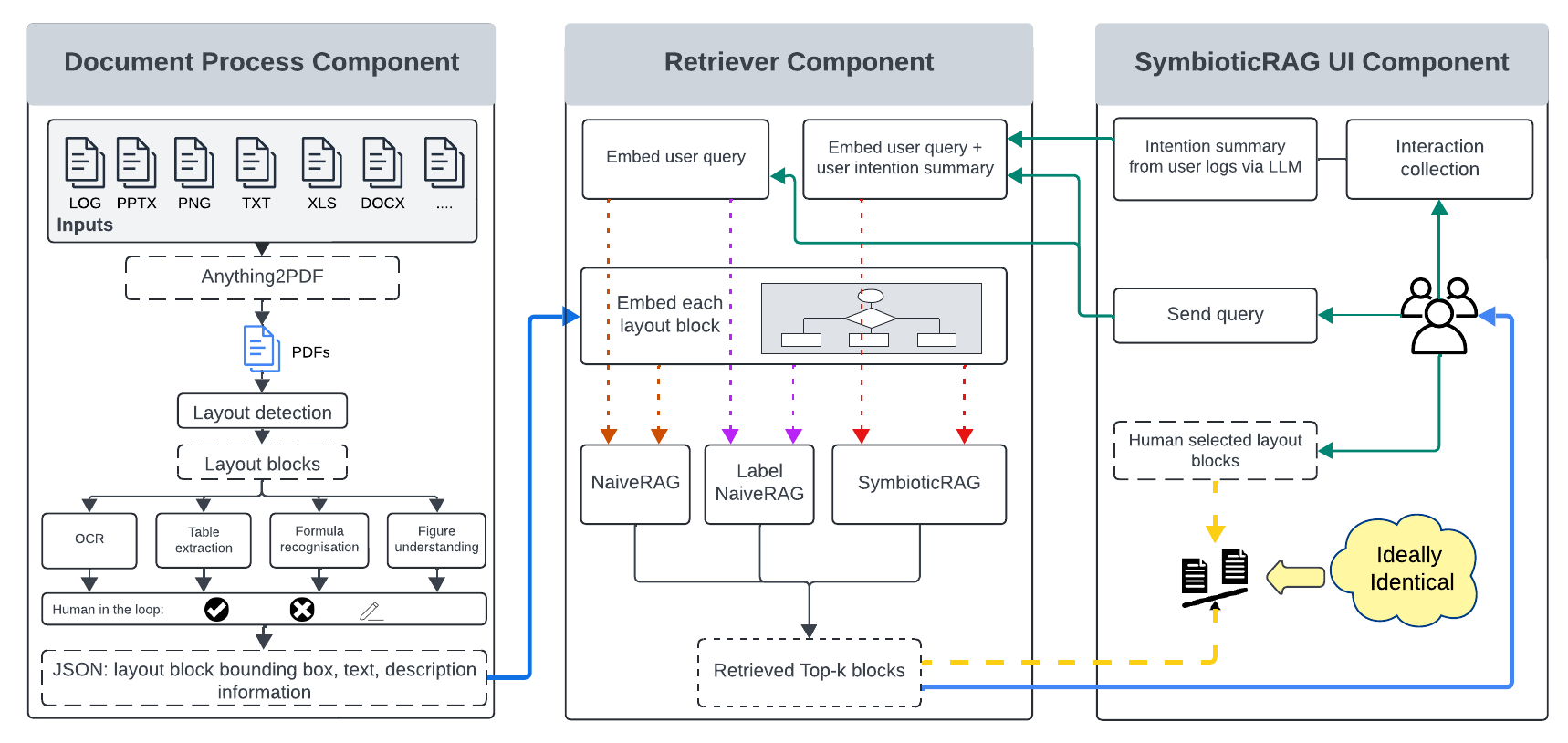}
\vspace{-1em}
\caption{The overview of SymbioticRAG system design}
\label{fig:system-design}
\vspace{-1em}
\end{figure*}

\begin{figure}[hbt]
    \centering
    \includegraphics[width=0.6\linewidth]{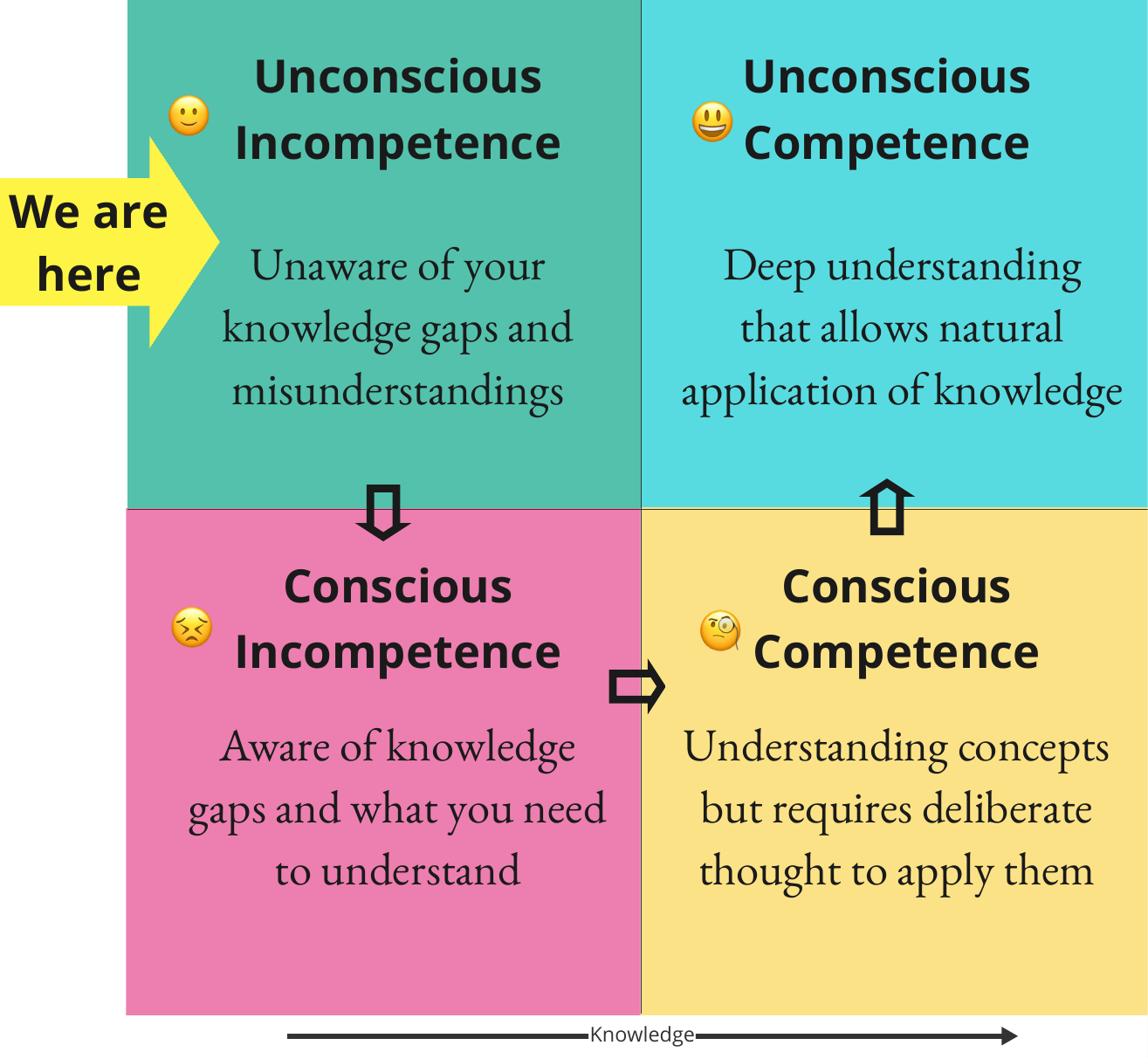}
      \vspace{-1em}
    \caption[Four stages of competence]{Four stages of competence proposed by Noel Burch (source: Wikipedia\protect\footnotemark). Current RAG systems effectively address \textbf{Conscious Incompetence}---stage where users can recognize knowledge gaps. However, they struggle with \textbf{Unconscious Incompetence}---users remain unaware of key knowledge deficits, necessitating more exploratory means to uncover these ``unknown unknowns.''}
    \label{fig:four-stage}
  \vspace{-0.5em}
\end{figure}

\textbf{Retrieval-Augmented Generation}~(RAG) addresses these issues by coupling LLMs with retrieval mechanisms that fetch relevant documents based on a user's query~\cite{craswell2022overview}. 
The grounding in external content helps reduce hallucinations and provides up-to-date information. 
Yet, deeper challenges remain. 
One concerns the distinction between \textbf{similarity} and \textbf{relevance}: retrieval components often return top-ranked chunks by embedding similarity, which may fail to capture tangential but crucial aspects of the query. 
Graph-based approaches such as GraphRAG~\cite{graphragmicrosoft} seek to introduce greater \textbf{diversity} into “relevance” by exploring knowledge graph structural features via community detection.
However, the ultimate \textbf{arbiter} is still the end users—they decide whether retrieved contents are relevant or not—underscoring that human should be at the centre of this retrieval process, as individuals often may have varying goals or perspectives even the query is the same.
Moreover, current RAG pipelines typically assume that users know what they need to ask, aligning well with situations in which knowledge gaps are consciously recognized~\cite{zhao2024retrievalaugmentedgenerationrag}, which illustrates that human centered design needs to be considered:
\begin{itemize}
    \item  \textbf{From unknown unknowns to knowledge discovery}: real-world learning often starts with ``Unknown Unknowns'' (or \textit{Unconscious Incompetence}), as shown in Figure~\ref{fig:four-stage}, where individuals do not yet realize the breadth of what they lack. 
For example., when a PhD student embarks on an unfamiliar topic like quantum computing, they may not even know what questions to pose. 
A simple query-response loop under RAG may suffice for direct, well-defined questions, but it falls short in supporting more human centered, exploratory, and iterative learning process through which novices actively engage with and gradually uncover the breadth of a domain.
    \item \textbf{Enhancing retrieval precision}: another limitation is the lack of a \textit{last-mile} correction mechanism, which allows users to refine retrieval when the content is close but not quite accurate.
\end{itemize}

\footnotetext{\url{https://en.wikipedia.org/wiki/Four_stages_of_competence}}

To address these challenges in document intelligence, we need to fundamentally rethink human-machine interaction. This leads us to propose a symbiotic framework - a system design principle inspired by biological symbiosis where two species live in close association and benefit from each other. In our context, humans and the RAG system act as two ``species'' that mutually influence and enhance each other's capabilities, eventually serving human needs.

The framework should integrates complementary strengths from existing paradigms: it preserves the successful interaction pattern of traditional search engines where users can retrieve and examine original documents, while leveraging LLMs' capabilities in fine-grained content retrieval and synthetic result generation. The symbiotic relationship operates bidirectionally: \textit{from system to human}, where provided information, explanations, and answers shape users' thought processes and knowledge acquisition as shown in Figure~\ref{fig:four-stage}; and \textit{from human to system}, where the system learns to understand and adapt to individual needs based on user behavior. This design specifically targets the limitations of current approaches: unlike traditional search engines that only support single-round document retrieval, our approach enables multi-round, context-aware document interactions to support exploratory learning. It also differs from current LLM-based systems such as ChatGPT or existing RAG systems, which primarily operate in a one-directional manner and lack comprehensive interactive behaviors between human and system. Recent developments like OpenAI's Canvas\footnote{\url{https://openai.com/index/introducing-canvas/}} demonstrate the industry's recognition of the importance of human-machine bidirectional interaction, though their current focus on post-generation editing has yet to extend to supporting user interaction during the external retrieval process itself.

We present \textbf{SymbioticRAG}, a human--LLM collaborative system designed to address the key limitations discussed above and enhance document intelligence. 
The solution comprises three core main components: \textbf{Document Processing}, \textbf{Retriever}, and \textbf{SymbioticRAG UI}.
\begin{itemize}
    \item  \textbf{Document Processing}: SymbioticRAG supports diverse document formats by first converting them into PDFs and processing each page as an image.
A layout detection module identifies bounding boxes for individual layout blocks, such as text blocks, titles, tables, figures, and formulas. Subsequently, these are processed by specialized modules for OCR, table extraction, figure understanding, and formula recognition. This process transforms the raw content into layout-aware representation, enabling fine-grained retrieval once the layout block chunks are embedded.
\end{itemize}

\begin{itemize}
\item \textbf{Retriever}: this module can accommodate various retrieval strategies, for example, the simplest semantic similarity-based retrieval over layout block embeddings, or integrate more advanced approaches as they emerge. This flexible architecture allows SymbioticRAG to adapt to different requirements and incorporate state-of-the-art retrieval methodologies.
The retrieved layout blocks will be presented within the context of their original documents, allowing users to verify relevance in situ. 
\end{itemize}

\begin{itemize}
\item \textbf{SymbioticRAG UI:} The interface enables users to explore the document space, refine, and iterate on retrieved layout blocks by clicking to include or remove specific ones. This interactive process allows precise curation of relevant layout blocks for downstream tasks such as report generation. It is natural and not onerous, which allows for subconsciously high quality user interaction data collection. 
\end{itemize}

We implemented \textbf{SymbioticRAG} and evaluated it in three scenarios: \textit{geological report exploration}, \textit{research literature review}, and \textit{education}. 
Our tests assessed its effectiveness and user satisfaction by comparing retrieved layout blocks from different retrieval strategies with those selected by users in multi-turn interactions. We also collected user feedback on interaction design and usability. Results show that \textbf{SymbioticRAG} outperforms traditional RAG systems, including ChatGPT and Claude, in user satisfaction and engagement. 
Its human-centered design enabled tasks previously impractical via current RAG systems, demonstrating its versatility across diverse domains.

\section{Motivation}\label{sec:motivation}

\paragraph{Retrieval-Augmented Generation~(RAG)}
LLM hallucination arises from misalignment between training data and reference sources~\cite{huang2024survey,llmhallucination}, often due to heuristic data collection or the generative nature of NLG tasks. As this issue remains unsolved, grounding model outputs in source documents has become essential, fueling the rapid development of RAG.
RAG retrieves the most \textbf{relevant} content for user queries. A basic approach ranks retrieved chunks using cosine similarity, equating \textbf{relevance} with \textbf{semantic similarity}. However, information retrieval~(IR) research distinguishes the two~\cite{irrelevance}, emphasizing that relevance also depends on task context, timeliness, and credibility, etc~\cite{userdefinedrelevance,similarity}. Bridging this gap remains an active research challenge~\cite{craswell2022overview}.
Recent RAG advancements address this issue through two main approaches. The first introduces more diverse features into \textbf{Relevance} by integrating graph structures, temporal information, and domain-specific metadata, etc. GraphRAG~\cite{graphragmicrosoft,graphRAGjianren} incorporates graph-based features to enhance retrieval, while HippoRAG~\cite{HippoRAG} leverages knowledge graphs and PageRank for filtering. 
The second approach shifts toward end-to-end models that directly learn relevance. For instance, Multi-Head RAG~\cite{multiheadrag} further refines relevance scoring using transformer-based attention mechanisms.
Dense Passage Retrieval (DPR)~\cite{dpr} optimizes dense embeddings using contrastive learning, while G-Retriever~\cite{he2024gretriever} constructs subgraphs and applies a Steiner tree strategy to compute quantifiable relevance values, integrating these metrics into LLMs via fine-tuning.
These developments raise a key question: as the user is the ultimate \textbf{arbiter} of relevance, should user behavior data be incorporated into model training? Recent studies explore this idea, with \cite{au2025personalizedgraphbasedretrievallarge,zerhoudi2024personaragenhancingretrievalaugmentedgeneration} leveraging user profiles for re-ranking and \cite{bai2024pistisragenhancingretrievalaugmentedgeneration} using feedback signals (e.g., dislikes, regenerations) to refine relevance scores.

\paragraph{Symbiotic interaction}
Symbiosis, a biological concept describing long-term mutual interaction between species, has inspired human-machine collaboration since at least 1960, when J.C.R. Licklider introduced \textbf{human-computer symbiosis} in ``Man-Computer Symbiosis''~\cite{mancomputer1960}. He proposed two key goals: integrating computers into early problem-solving stages for collaborative question formulation and enabling real-time human-computer interaction for immediate feedback and iteration.
Recent advancements in LLMs have made real-time interaction more feasible. For example, OpenAI's GPT-4o achieves a latency of approximately 196 milliseconds per token\footnote{\url{https://openai.com/index/hello-gpt-4o/}}, approaching the threshold of human real-time perception. However, the challenge remains in designing effective human-machine collaboration mechanisms that guide users from ``unconscious incompetence'' to ``conscious incompetence'' in question formulation.
Despite growing recognition of symbiotic human-machine systems as an important direction~\cite{SymbioticDiscussion1,SymbioticApproach,islami2024human,10.1145/3661167.3661223,lin2024progress,abbass2024future}, practical implementations remain in early stages. In document intelligence, Symbiotic Recommendations~\cite{SymbioticRecommendations} injects user profiles into prompts via prompt engineering, similar to personalization techniques in RAG~\cite{au2025personalizedgraphbasedretrievallarge,zerhoudi2024personaragenhancingretrievalaugmentedgeneration,bai2024pistisragenhancingretrievalaugmentedgeneration}. However, deeper, bidirectional adaptations between humans and machines are still largely unexplored.

To bridge this gap, we propose \textbf{SymbioticRAG}, which enables users to directly select retrieved content and later incorporates user interactions into an end-to-end \textbf{Relevance} model. By continuously learning from user behaviors, the system aims to establish a positive feedback loop, evolving into a personalized, adaptive agent that tailors retrieval and generation to individual needs. This is the vision behind \textbf{SymbioticRAG}, which marks a step toward true human-machine symbiosis.

\section{System design}

The fundamental design philosophy of RAG places humans at the center. 
1).~Instead of approximating relevance metrics to model user intent, we advocate restoring decision-making authority to humans.  
2).~Human-selected content for each query then serves as training data, enabling the system to adapt to user intentions through explicit selection and interaction patterns.  
3).~To bridge the gap between ``Unconscious Incompetence'' and ``Conscious Incompetence'' in query formulation, we propose maintaining shared access to source documents for both humans and LLMs. This allows users to navigate the document space independently, enhancing their understanding and awareness of available information.  
Through this approach, we establish a symbiotic relationship where machines gradually refine their understanding of human intent, while humans gain diverse perspectives, fostering collaborative exploration of complex information seeking tasks.




\begin{figure}
    \centering
  
    \includegraphics[width=0.72\linewidth]{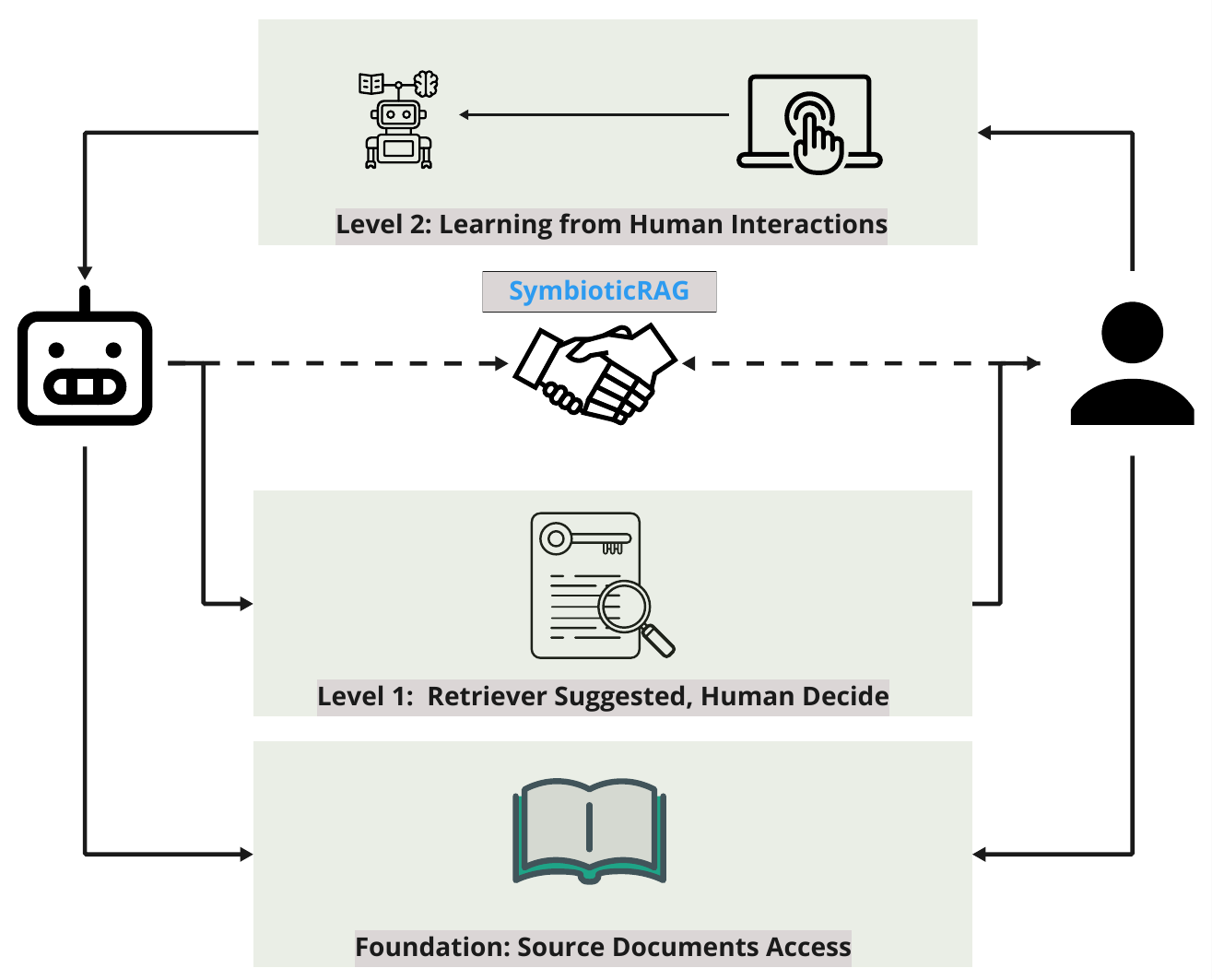}
    \caption{SymbioticRAG concept illustration}
    \label{fig:symbiotic-concept}
    \vspace{-1em}
\end{figure}

We define three key characteristics of a \textbf{SymbioticRAG} system, which is illustrated in Figure~\ref{fig:symbiotic-concept}:
(1)~The foundational feature ensures direct document access and readability, allowing humans to independently explore source content.  
(2)~\textbf{Level 1} establishes user-driven retrieval, where humans, with reference to machine-retrieved content, actively select relevant information to augment prompts for answer generation. 
(3)~\textbf{Level 2} continuously refines retrieval models by learning from human interactions and selected content, enabling personalized content retrieval.  
Most existing RAG systems, particularly those handling large document collections, lack direct human access and content selection mechanisms—both essential for achieving true symbiosis. 
This paper focuses on implementing the foundational feature and Level 1, experimenting and laying the groundwork for future Level 2 advancement.


As shown in Figure~\ref{fig:system-design}, our system consists of three main components:  
1)~\textbf{Document Processing}: This pipeline standardizes document formats into PDF, detects layout blocks with bounding boxes, and applies OCR, table extraction, formula recognition, and figure understanding to convert documents into LLM-digestible text.  
2)~\textbf{Retriever}: This extensible module supports multiple retrieval strategies, including semantic search. It embeds and indexes layout blocks in a vector database, retrieving relevant blocks to augment LLM prompts.  
3)~\textbf{SymbioticRAG UI}: Users interact with retrieved content by exploring matched layout blocks within source documents, understanding their context, exploring through the source documents and manually selecting blocks for further conversations or answer regeneration.  


\subsection{Document processing}

\begin{figure*}[hbt]
\centering
\begin{subfigure}[b]{0.29\linewidth}
\centering
\includegraphics[width=\linewidth]{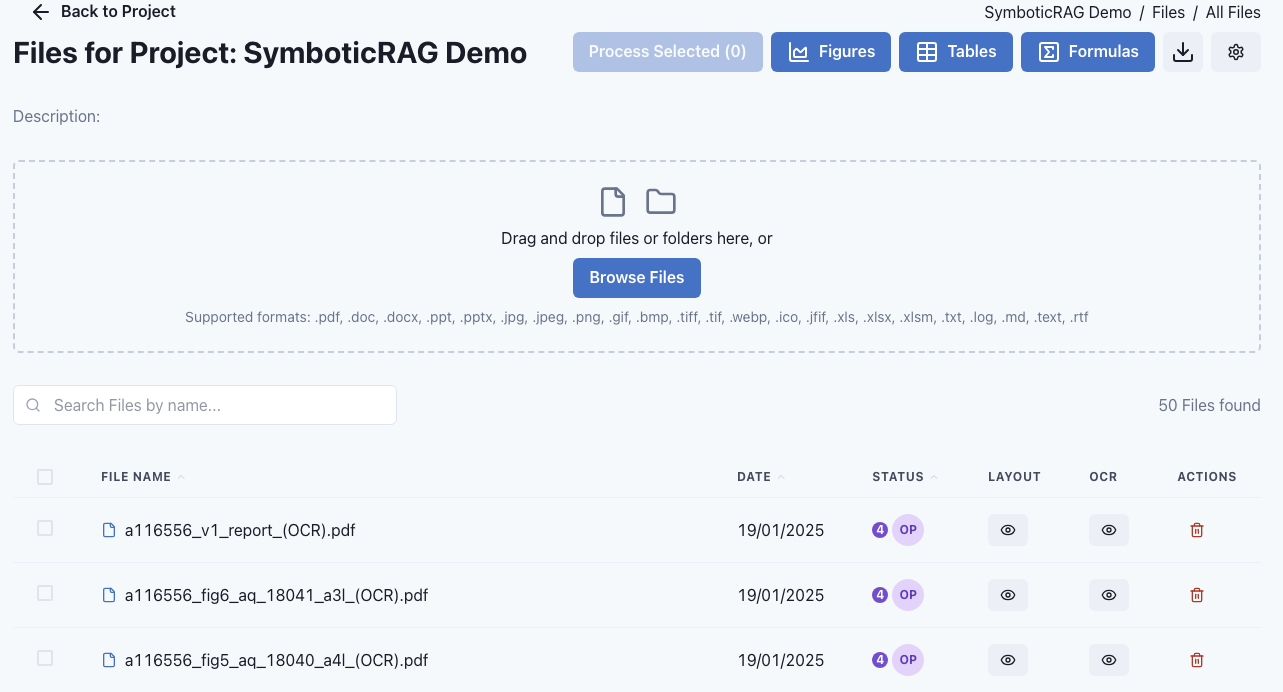}
\caption{Document processing pipeline dashboard for file upload, processing initiation and progress monitoring.}
\end{subfigure}
\hfill
\begin{subfigure}[b]{0.29\linewidth}
\centering
\includegraphics[width=\linewidth]{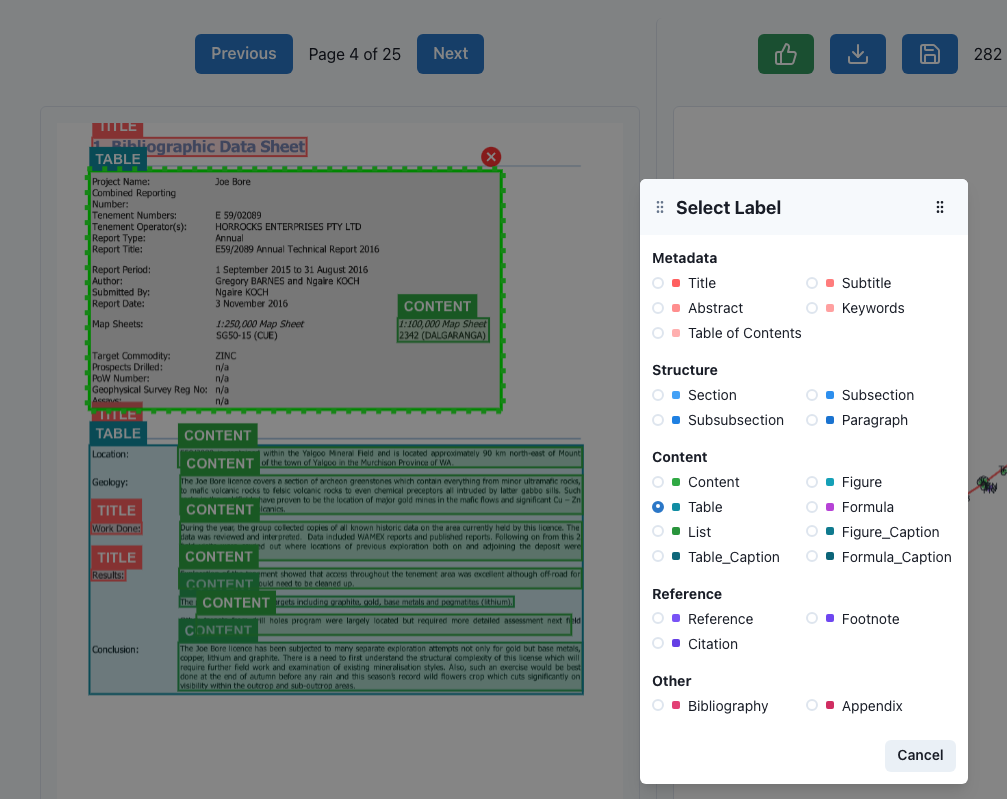}
\caption{Layout validation interface where users can review and edit detected layout block, including block reclassification, addition, removal and boundary adjustments.}
\end{subfigure}
\hfill
\begin{subfigure}[b]{0.29\linewidth}
\centering
\includegraphics[width=\linewidth]{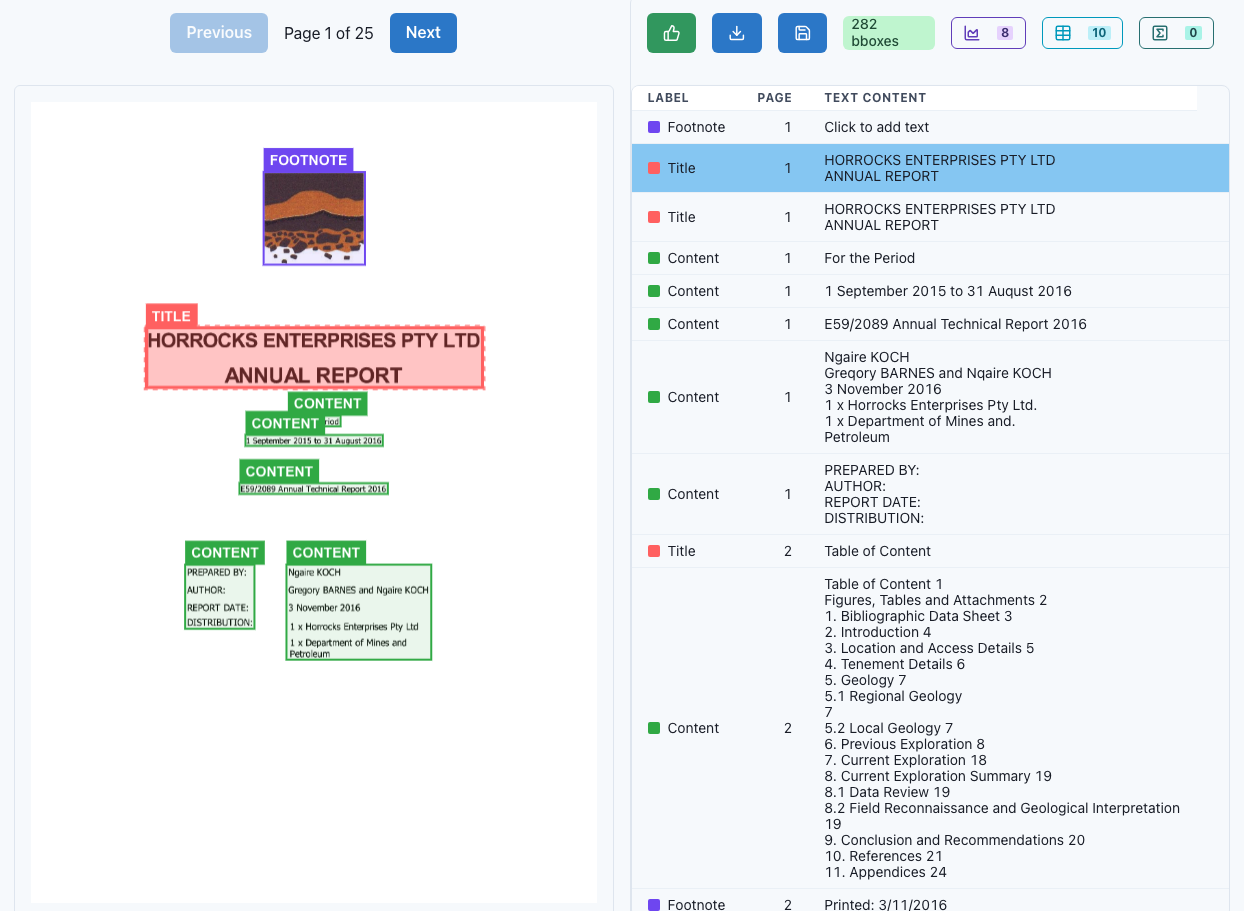}
\caption{OCR validation interface for reviewing and correcting text recognition results, particularly useful for handwritten text where OCR models struggle.}
\end{subfigure}
\vspace{0.2cm}
\begin{subfigure}[b]{0.29\linewidth}
\centering
\includegraphics[width=\linewidth]{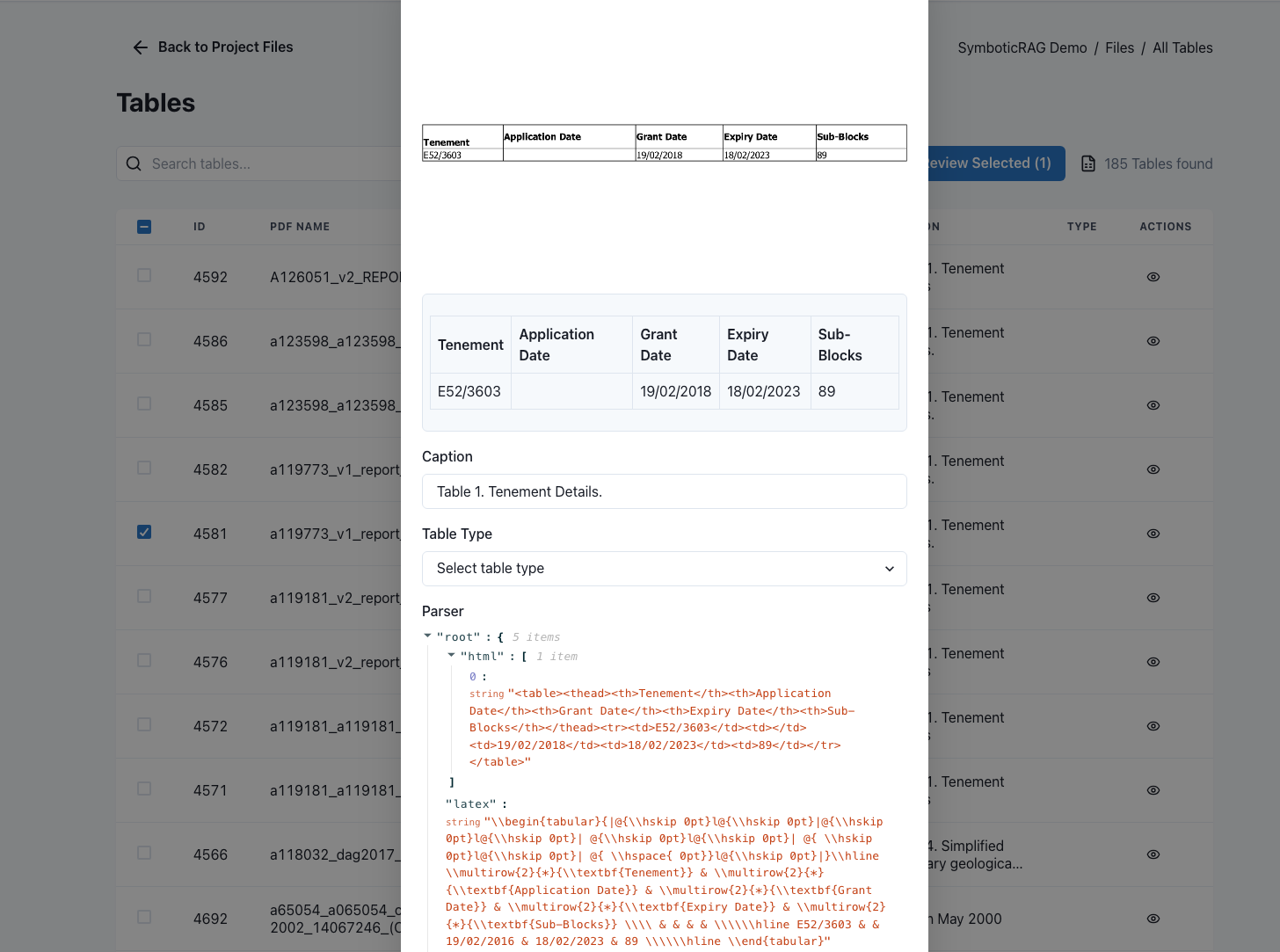}
\caption{Table validation interface that presents extracted table outputs in JSON viewer, allowing users to review, correct or add extra contents. The interface also supports batch review to increase efficiency.}
\end{subfigure}
\hfill
\begin{subfigure}[b]{0.29\linewidth}
\centering
\includegraphics[width=\linewidth]{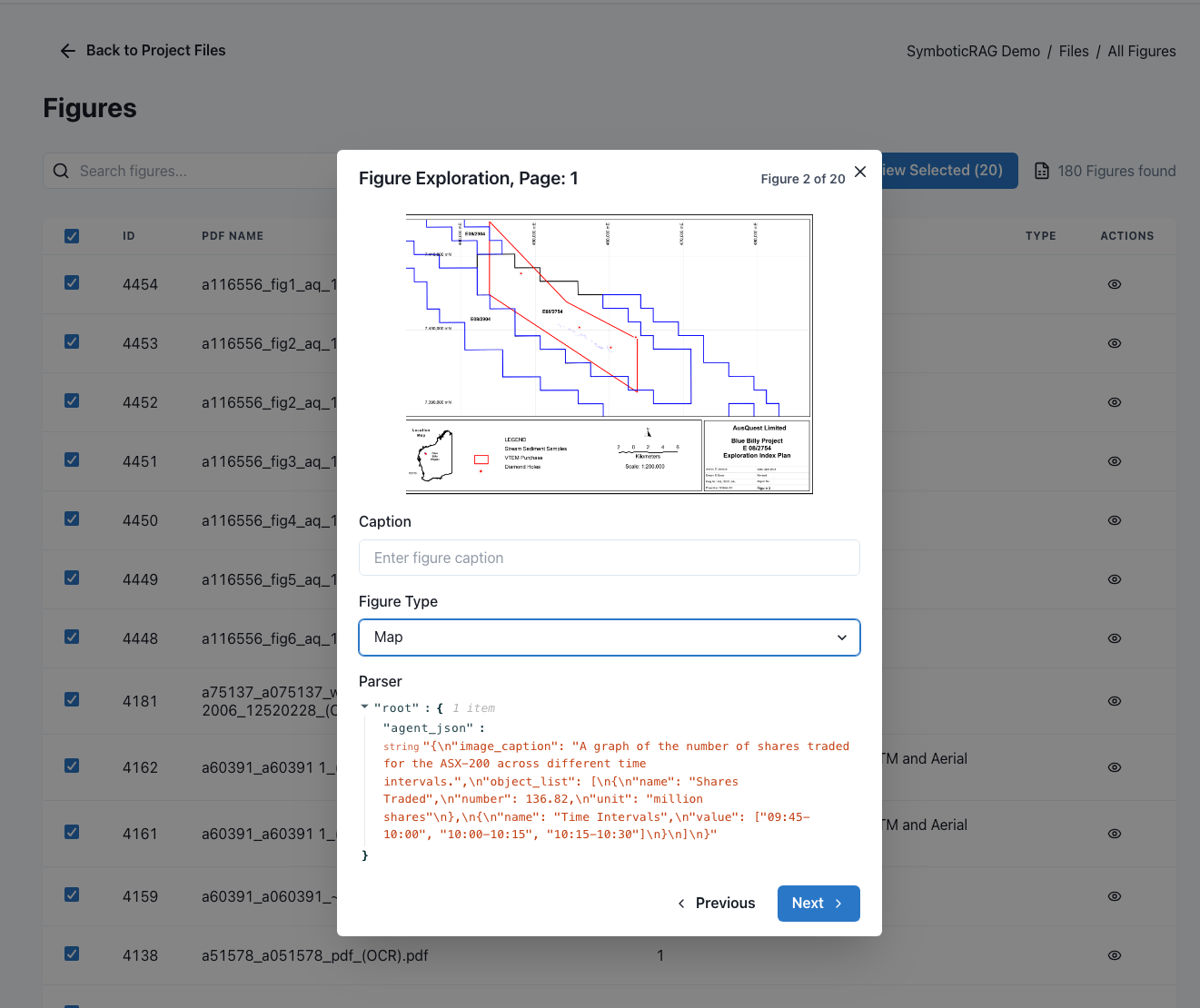}
\caption{Figure validation interface where users can verify, modify and add figure descriptions. Users can also review and update figure captions and types.}
\end{subfigure}
\hfill
\begin{subfigure}[b]{0.29\linewidth}
\centering
\includegraphics[width=\linewidth]{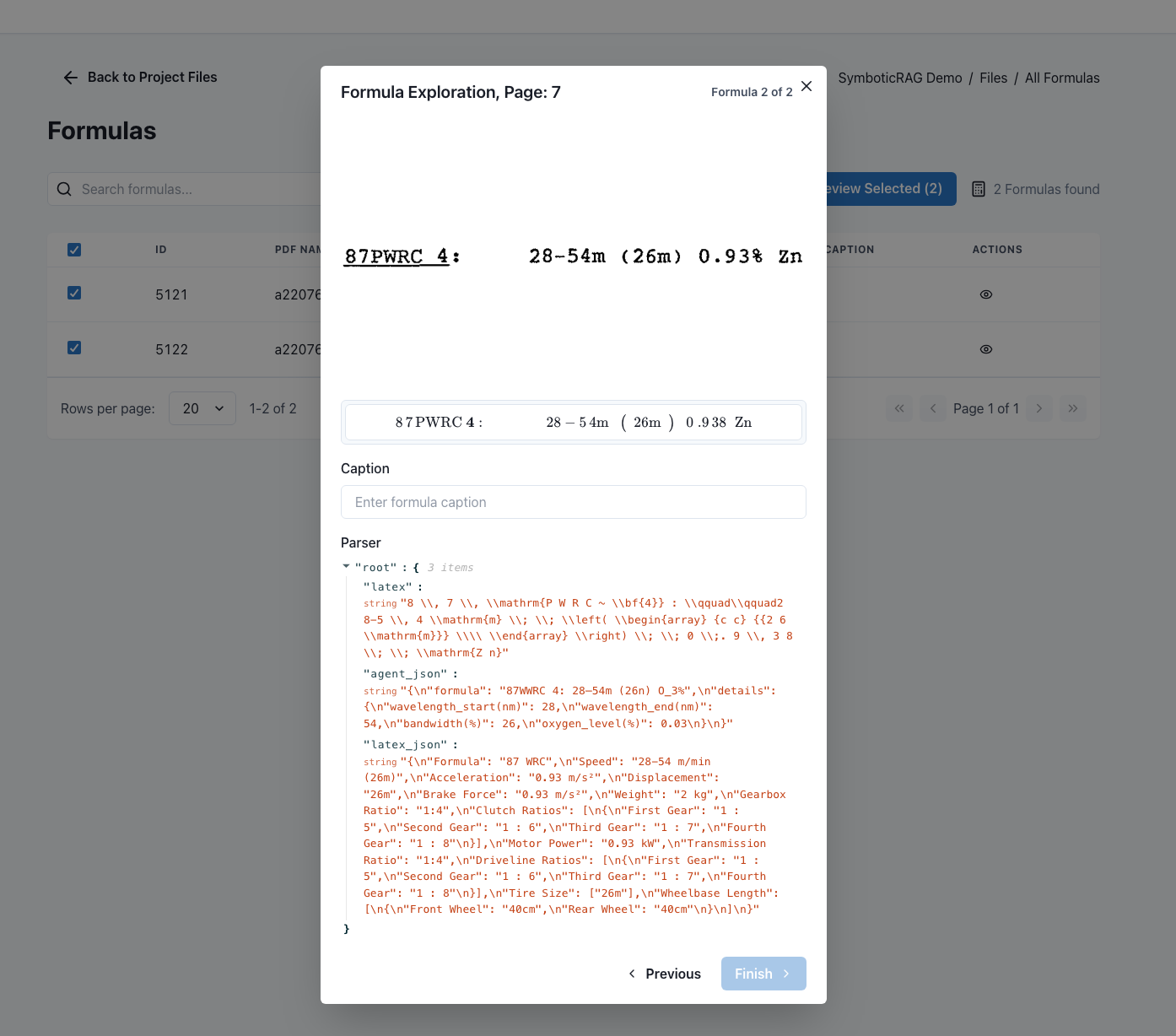}
\caption{Formula validation interface for reviewing and correcting mathematical formula extraction results in latex format and descriptive information about the formula.}
\end{subfigure}
\vspace{-1em}
\caption{Human-on-the-loop validation interfaces for the document processing pipeline, supporting comprehensive validation of layout analysis, OCR, table extraction, figure processing and mathematical formula recognition and understanding.}
\label{fig:validation-interface}

\end{figure*}

Despite the variety of document formats (e.g., Word, Excel, PDF, images), documents fall into two fundamental categories~\cite{docs2kg}: \textbf{digitally native} and \textbf{image-based}. Digitally native documents contain machine-readable text that can be reliably extracted via rule-based conversion, such as plain text files and exported PDFs. In contrast, image-based documents, including historical manuscripts and photographs, pose significant extraction challenges.

Recent research on image-based document processing begins with layout detection~\cite{zhao2024doclayoutyoloenhancingdocumentlayout}, identifying and classifying layout blocks (e.g., tables, formulas, figures, titles, content blocks). These are then processed by specialized models: OCR~\cite{du2020ppocrpracticalultralightweight} for text, dedicated models for table extraction, formula recognition, and figure understanding. While OCR and basic table/formula conversion have improved, complex table extraction, figure understanding, and handwritten formula recognition remain difficult~\cite{du2020ppocrpracticalultralightweight,xia2024docgenome,TRUONG2024110531}. 
Popular open-source systems such as MinerU~\cite{mineru} and DocLing~\cite{Docling} employ dual-path processing to effectively convert PDFs into markdown or JSON.
However, these transformations merge and re-segment layout blocks, losing precise positional information relative to the source documents, which is essential for our system design.

To address this, we implement a unified document processing pipeline that standardizes all input documents to \textbf{PDF format} before applying image-based techniques. This approach simplifies processing while preserving precise source attribution through layout block bounding boxes. Our design emphasizes simplicity, robustness, and extensibility, prioritizing accurate document source tracking.\\
\noindent\textbf{Layout Detection.} Our pipeline starts with DocLayout-YOLO~\cite{zhao2024doclayoutyoloenhancingdocumentlayout}, identifying bounding boxes and semantic classes (titles, content, tables, figures, etc.). Detected blocks are then processed by specialized modules: \\
\textbf{OCR.} We employ PaddleOCR~\footnote{\url{https://paddlepaddle.github.io/PaddleOCR/latest/en/index.html}} for its robust performance, multi-language support, and stability. \\
\textbf{Table.} The optimal output format for table extraction in RAG systems remains open. We explore three approaches: StructEqTable~\cite{xia2024docgenome} for LaTeX generation, Pix2Text~\cite{PIX2TEXTTABLE} for HTML, and a visual LLM method that produces structured JSON directly from table images to describe the table content. \\
\textbf{Formula.} Although Pix2Text~\cite{mathformula} reliably extracts formulas as LaTeX, mathematical expressions often contain domain-specific notations and complex semantics. To enhance downstream applications, we augment the LaTeX output with semantic descriptions generated by visual LLMs (e.g., llama3.2-vision\footnote{\url{https://huggingface.co/meta-llama/Llama-3.2-11B-Vision}}).\\
\textbf{Figure.} Figure understanding is domain-dependent. While visual language models can interpret general-purpose illustrations, specialized models perform better in fields like medicine~\cite{dhote2023surveyfigureclassificationtechniques} and science~\cite{shi2024mattersintegrityverificationscientific}. 
However, effectiveness varies with domain complexity. As a first step, we use visual LLMs to produce descriptive text summaries, similar to our formula approach, while acknowledging that more domain-specific solutions will be required.
\textbf{Human on the loop validation.} 
The immaturity of certain document processing techniques—especially for figures and formulas—poses significant challenges in maintaining high-quality data from diverse document formats for downstream retrieval tasks. To address this while advancing specialized methods (e.g., table extraction, figure and formula understanding), we introduce a \textit{human-on-the-loop} validation interface that supports human review at each processing step. Unlike traditional manual annotation approaches, our system focuses on validating model-generated outputs, balancing reduced human effort with improved data quality and benefiting iterative model refinement.

Our approach is especially beneficial in domains where high-quality training data is scarce, as it positions human reviewers as both overseers and guides of the pipeline. Consequently, it yields high-quality structured data for downstream tasks and fosters model development in underrepresented fields, further highlighting its human-centered nature.
Figure~\ref{fig:validation-interface} illustrates the validation system. 


\begin{figure*}[hbt]
    \centering
    \includegraphics[width=0.99\linewidth]{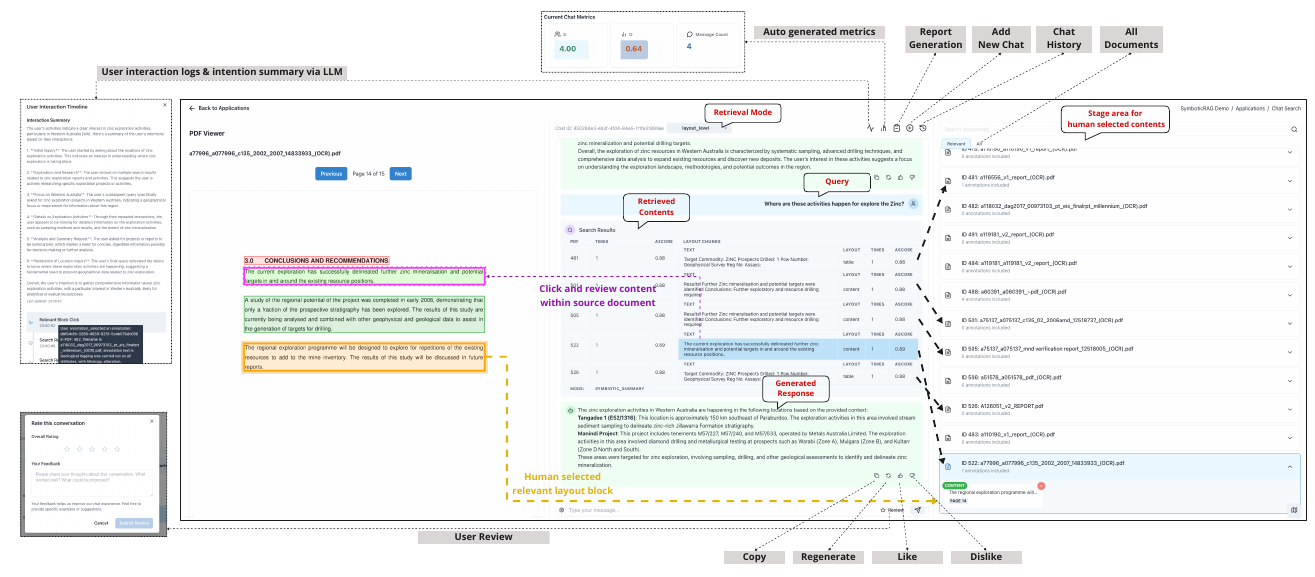}
    \caption{SymbioticRAG UI demonstration example}
    \label{fig:SynbioticRAGUI}
    \vspace{-1em}
\end{figure*}

\subsection{Retriever}
Although new, more effective retrieval methods are emerging, our focus is on improving retrieval through end-user behavior. To accommodate future advances, we designed an \textbf{extensible retrieval module} that can easily integrate newer retrievers. For initial testing, we implemented two baseline methods: \textit{Na\"{\i}veRAG}, which performs semantic similarity search by embedding each layout block (including tables, formulas, and figures) using the E5 model~\cite{wang2024textembeddingsweaklysupervisedcontrastive} and retrieving the top-$k$ blocks; and \textit{LabelNa\"{\i}veRAG}, which retrieves the top-$k$ blocks separately for each block type and then merges the results for overall top-$k$ layout blocks. 

\subsection{SymbioticRAG UI}
The SymbioticRAG user interface~(UI) consists of three primary components, as illustrated in Figure~\ref{fig:SynbioticRAGUI}. The left side \textbf{PDF Viewer} displays annotated source documents and enables \textit{select} and \textit{deselect} interactions, while the middle \textbf{Chat} component facilitates multi-turn conversations for query submission, retrieval results exploration, and AI-generated responses review. The third component, a \textbf{Staging} area, maintains a collection of human-selected relevant layout blocks.

Following document processing, the conversation starts with user queries (highlighted in blue). 
A defined retriever identifies the top-k matched layout blocks (currently k=5), presenting the searched results as table format in the chat interface (highlighted in purple). 
Meanwhile, the augmented query is processed by an LLM (currently GPT-4o) to generate a response (displayed in green).

Users can navigate to the source document by clicking any search result, which opens the selected layout block in its original context. For example, in the demo figure, selecting the first result jumps to page 14 (in Fig~\ref{fig:SynbioticRAGUI}), highlighting the relevant block with a purple dashed rectangle in the PDF Viewer. Users can explore surrounding content, assess relevance, and toggle block selection—selected blocks are marked with a yellow double solid line, presented and grouped by source document in the staging area. They can review multiple \textit{Retriever}-suggested blocks across documents to refine understanding before follow-up queries. Clicking the regeneration button incorporates human-selected blocks into augmented prompts for updated responses. To mitigate cases where \textit{the real relevant} documents are not present in the retrieved layout blocks, users can access all documents via the \textit{All} tab. The interface also includes chat management features to start new conversations and review chat histories.

During user interactions, we record engagement activities such as sending queries, clicking search results, selecting/deselecting blocks, navigating pages, manually adding documents, and liking/disliking or regenerating responses. These logs will inform our development of \textbf{SymbioticRAG Level 2}. We currently feed them into an LLM to generate a user-intention summary~(in Fig~\ref{fig:SynbioticRAGUI}), which is then concatenated with the query for semantic similarity search. We label this retriever strategy: \textit{SymbioticRAG}, experimenting how future \textbf{SymbioticRAG Level 2} can integrate user feedback.
We also tried to concatenate user logs directly with query and then generate a query embedding, however, due to the raw text match from the content, the semantic similarity retrieved content will converge to specific contents which already exists inside the user logs.
Due to this limitation, we excluded this approach from our comparative analysis.

\begin{figure}[hbt]
    \centering
    \includegraphics[width=0.99\linewidth]{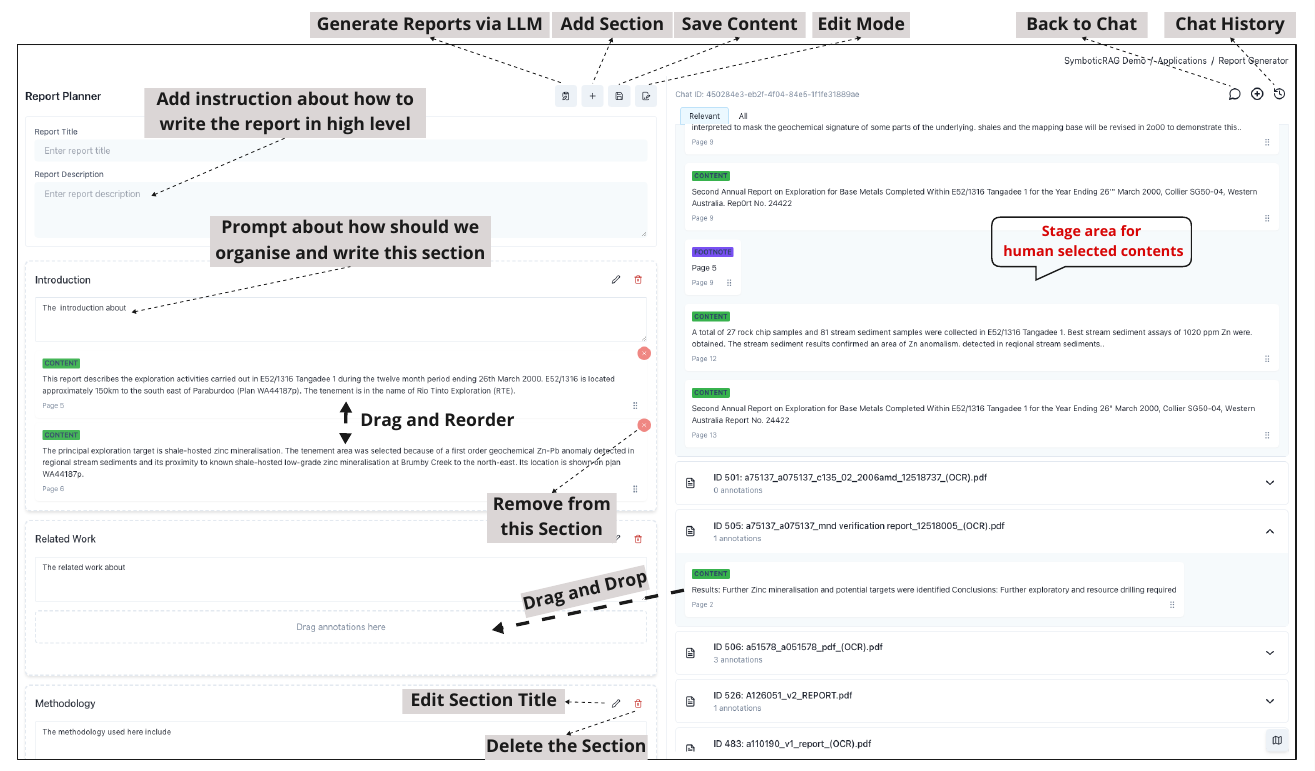}
    \caption{Report generation interface example for SymbioticRAG}
    \label{fig:report-generation}
    \vspace{-1.5em}
\end{figure}

\paragraph{Report generation} 
Report writing often demands substantial time to gather and organize evidence from multiple sources. Our system addresses this challenge via a staging area that allows users to collect and verify human-selected layout blocks through interactive conversations. As shown in Figure~\ref{fig:report-generation}, we provide a dedicated report generation interface that leverages these curated blocks. 
Users can outline their report structure, drag and drop relevant blocks into specific sections, and supply writing instructions for each component. The system then employs an LLM to generate a draft, with options for direct editing and exporting to Word format. This approach streamlines the process of collecting, verifying, and organizing evidence from multiple sources, addressing a key gap in traditional RAG systems and enhancing evidence-based report writing. High resolution of Figure~\ref{fig:SynbioticRAGUI} and Figure~\ref{fig:report-generation} will be included in the supplementary materials.

\section{Evaluation and Results}
We employ both quantitative metrics and qualitative feedback to comprehensively assess our Level 1 implementation and Level 2 experimental exploration of the proposed SymbioticRAG system.

\paragraph{Evaluation Scenarios}
We tested three distinct scenarios:
(i)~a \textit{literature review} scenario involving 30 academic papers related to RAG for systematic reviews,
(ii)~a \textit{geological exploration} scenario analyzing 30 zinc-focused reports reflecting typical tasks in geological departments, and
(iii)~an \textit{education} scenario incorporating ``Full Stack Developer'' Unit materials to support student learning.
\paragraph{Participants}
Three independent evaluators were assigned to each scenario, each conducting five conversation sessions per retrieval strategy~(\textit{Na\"{\i}veRAG}, \textit{LabelNa\"{\i}veRAG}, and \textit{SymbioticRAG}), followed by quantitative satisfaction ratings and post-experiment interviews. For the \textit{literature review} scenario, the evaluators were three computer science researchers; for the \textit{geological report} scenario, three experienced geologists; and for the \textit{education} scenario, three undergraduate students.
Identical questions were asked across retrieval strategies, with evaluators randomly alternating between strategies across sessions to mitigate learning bias, though this cannot be fully eliminated.

\paragraph{Evaluation metrics} 
We used two outcome-oriented metrics to assess system effectiveness. The first metric captures the layout blocks selected by users, representing the ``true relevant'' content for each conversation. The second is user satisfaction, rated on a 5-point scale (with 5 indicating ``very satisfied'').
To quantify alignment between user and retriever selections, we define the human-retriever distance $D$ as:
\begin{equation}
D = 1 - \frac{|H \cap R|}{|H \cup R|},
\end{equation}
where $H$ and $R$ are the sets of blocks selected by humans and the retriever, respectively. This metric ranges from 0 (perfect alignment) to 1 (complete divergence). We also compute a satisfaction metric $S$ as the mean user satisfaction rating, reflecting overall usability and effectiveness.


\paragraph{Results}

\begin{table}[H]
\vspace{-1em}
\caption{Comparative Evaluation Results Across Different Scenarios}
\vspace{-0.8em}
\renewcommand{\arraystretch}{1.5}
\resizebox{\linewidth}{!}{%
\begin{tabular}{llccc}
\hline
\addlinespace[3pt]
Strategy & Metric & Literature Review & Geological Reports & Education \\
\addlinespace[1.5pt]
\hline
\addlinespace[3pt]
\multirow{2}{*}{\textbf{Na\"{\i}veRAG}} 
& Human-Retriever Distance ($D$, 0-1) ↓ & 0.85 & 0.92 & 0.88 \\[6pt]
& User Satisfaction ($S$, 1-5) ↑ & 2.47 & 2.13 & 1.80 \\
\addlinespace[1.5pt]
\hline
\addlinespace[3pt]
\multirow{2}{*}{\textbf{LabelNa\"{\i}veRAG}} 
& Human-Retriever Distance ($D$, 0-1) ↓ & 0.78 & 0.83 & 0.81 \\[6pt]
& User Satisfaction ($S$, 1-5) ↑ & 3.13 & 2.93 & 2.67 \\
\addlinespace[1.5pt]
\hline
\addlinespace[3pt]
\multirow{2}{*}{\textbf{SymbioticRAG}} 
& Human-Retriever Distance ($D$, 0-1) ↓ & 0.52 & 0.61 & 0.58 \\[6pt]
& User Satisfaction ($S$, 1-5) ↑ & 4.13 & 3.93 & 3.67 \\
\addlinespace[1.5pt]
\hline
\end{tabular}}
\vspace{-1em}
\label{tab:evaluation_results}
\end{table}

Table \ref{tab:evaluation_results} presents the quantitative results of our evaluation across three scenarios. The Human-Retriever Distance metric ($D$) results reveal a significant challenge in aligning retriever outputs with human information needs, with \textit{Na\"{\i}veRAG} showing consistently high distances (0.85-0.92). This supports our hypothesis that users in unknown-unknown states often struggle to articulate their precise information needs. The introduction of \textit{LabelNa\"{\i}veRAG} showed modest improvements, reducing distances to 0.78-0.83 through increased retrieval diversity. Most notably, \textit{SymbioticRAG} demonstrated substantial gains, achieving distances of 0.52-0.61, suggesting that augmented queries with user interaction summary better capture user semantic intent. User satisfaction scores correlated with reduced retriever distances, with \textit{SymbioticRAG} achieving the highest ratings (3.67-4.13). Literature review and geological scenarios showed marginally better performance, likely due to participants' domain expertise. Post-experiment interviews revealed two critical features driving user satisfaction: the ability to examine source documents was described as ``game-changing,'' while the capacity to select and incorporate layout blocks into prompts addressed a key limitation in existing LLM interfaces.

\begin{figure}
    \vspace{-0.5em}
    \centering
    \includegraphics[width=0.9\linewidth]{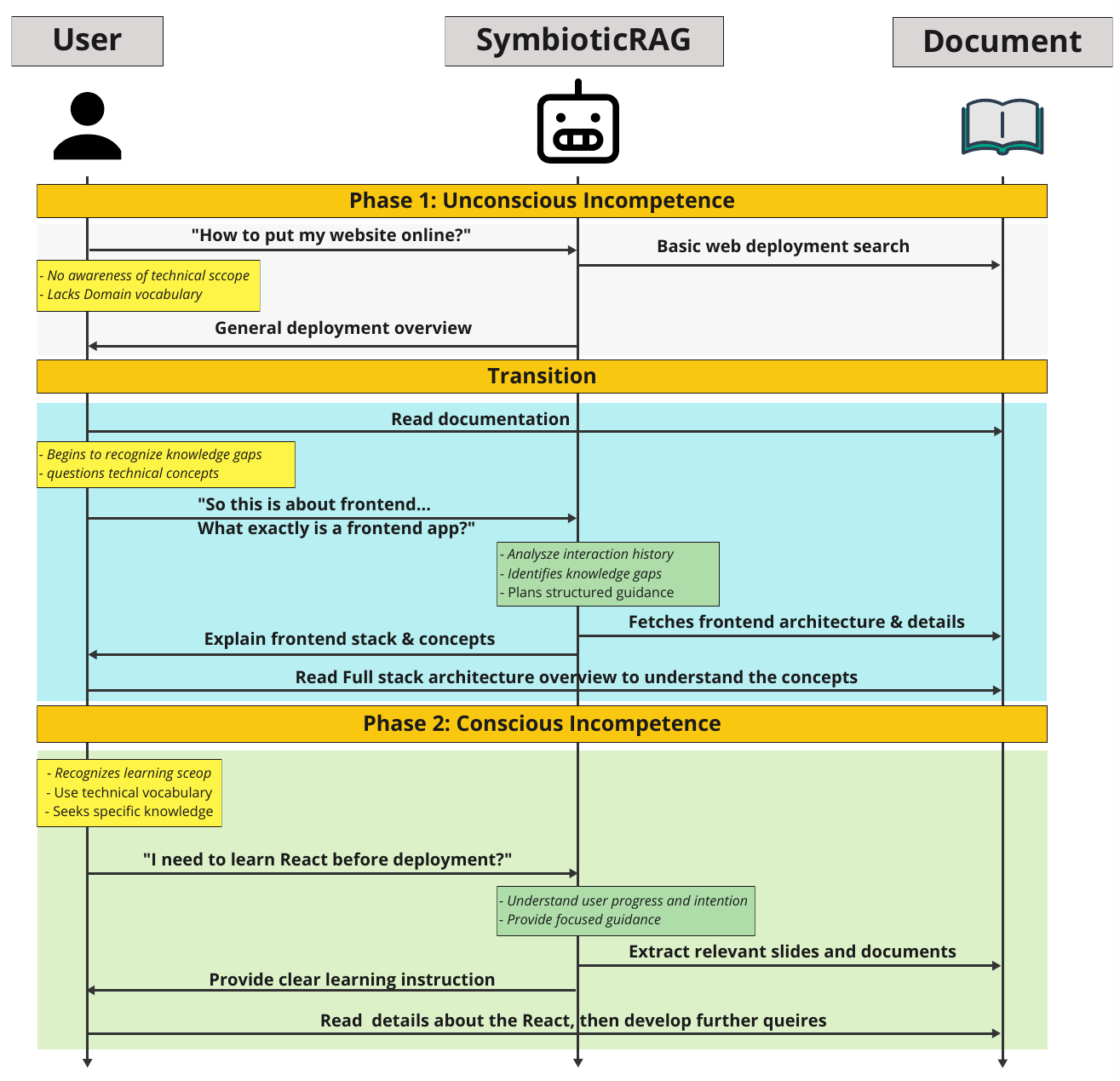}
    \caption{Case Study in \textit{education} scenario~(Full Stack Developer Unit) with \textit{SymbioticRAG} retriever.}
    \label{fig:casestudy}
    \vspace{-1.5em}
\end{figure}

\paragraph{Case study}
We analyzed a user's conversation~(in Figure~\ref{fig:casestudy}) in the \textit{education} scenario using \textit{SymbioticRAG} retriever. The interaction history reveals three phases: from initial ``Unconscious Incompetence'' with basic queries, through a ``Transition'' phase with emerging awareness, to ``Conscious Incompetence'' where users demonstrate proper technical vocabulary. This progression validates that our \textit{SymbioticRAG} design effectively positions users at the center, supporting their natural development of domain understanding.

\section{Conclusion}

We presented \textbf{SymbioticRAG}, a novel framework that fundamentally reimagines RAG systems through a human-centered lens. Our work addresses two critical challenges in current RAG systems: the inherently human nature of relevance determination and users' struggle with \textbf{Unconscious Incompetence} when formulating queries in unfamiliar domains. Our framework introduces a two-tiered approach: Level 1 enables direct human curation of retrieval content through interactive document access, while Level 2 aims to develop personalized retrieval models based on user interactions. We successfully implemented Level 1 with three key components: a comprehensive document processing pipeline, an extensible retriever module, and an interactive UI that facilitates user engagement while collecting valuable interaction data. To maintain high-quality data preparation, we developed a human-on-the-loop validation interface that improves pipeline output while advancing research in specialized extraction tasks.

Our evaluation across three distinct scenarios demonstrated the effectiveness of our approach. More importantly, our attempt at Level 2 implementation through \textit{SymbioticRAG}, which augments queries with summaries of user interactions, showed promising results with significantly improved retrieval relevance and user satisfaction scores. To facilitate broader research and further advancement of \textbf{SymbioticRAG} Level 2, we will make our system openly and freely accessible to the research community upon paper acceptation.

\bibliography{latex/ref}

\end{document}